\documentclass[12pt,a4paper]{article}
\usepackage{amsmath}
\usepackage{graphicx}
\usepackage{times}
\begin{document}
\begin{titlepage}
\title{ The new elastic scattering measurements of TOTEM --- are there hints for  asymptotics? }
\author{ Sergey M. Troshin, Nikolai E. Tyurin\\[1ex]
\small   NRC ``Kurchatov Institute''--IHEP\\
\small   Protvino, 142281, Russian Federation\\
\small   Sergey.Troshin@ihep.ru}
\normalsize
\date{}
\maketitle

\begin{abstract}
We point out to another indication of the black-disk limit exceeding in hadron interactions found in the recent impact parameter analysis performed by the TOTEM Collaboration  at $\sqrt{s}$=8 TeV and emphasize  that this observation might be interpreted as a confirmation of the reflective scattering mode appearance at the LHC energies. 
\end{abstract}
\end{titlepage}
\setcounter{page}{2}

The geometrical picture of hadron interactions is often based on the impact--parameter dependence of the inelastic overlap function. However, such approach is not quite complete, i.e. elastic and total overlap functions are to be considered in line. On the ground of such consideration
a slow gradual transition to the emerging at the LHC picture,  where the interaction region starts to become reflective at the center ($b=0$) and simultaneously becomes relatively edgier, larger and black at its periphery (we are using an acronym  REL to denote this picture) has been discussed in \cite{mpla16}, where the references for the earlier papers can be found. The transition to this mode  seems to be observed by the TOTEM experiment under the measurements of the $d\sigma/dt$ in elastic $pp$--scattering. This is based on the analysis   of the impact parameter dependences of the overlap functions performed in \cite{alkin}. Those overlap functions enter unitarity relation: 
\begin{equation}\label{un}
\mbox{Im}f(s,b)=h_{el}(s,b)+h_{inel}(s,b).
\end{equation}
In Eq. (\ref{un}) the function $f(s,b)$ is the elastic scattering amplitude in the impact parameter representation while $h_{el}(s,b)$  is the elastic overlap function, \[h_{el}(s,b)={|f(s,b)|}^2. \] 
The inelastic overlap function $h_{inel}(s,b)$ corresponds to the total contribution of all the inelastic processes.
From Eq. (\ref{un}) the following inequality for the real part of the scattering amplitude $\mbox{Re}f(s,b)$  is obtained \cite{phlt}:
\begin{equation}\label{re}
-\frac{1}{2}\sqrt{1-4h_{inel}(s,b)}\leq \mbox{Re}f(s,b)\leq \frac{1}{2}\sqrt{1-4h_{inel}(s,b)}.
\end{equation}

It should be noted that Eq. (\ref{un}) is an approximate one due to the kinematical constraints existing at finite energies.
It is valid with an accuracy of ${\cal{O}} (1/s)$ \cite{gold} being a result of the Fourier-Bessel transform of the unitarity equation written in  $s$ and $t$ variables:
\begin{equation}\label{unt}
\mbox{Im}F(s,t)=H_{el}(s,t)+H_{inel}(s,t).
\end{equation}

Considering the limit $s\to \infty$,  the  question on the limiting value for the scattering amplitude can be posed,  is it the black disk limit or the unitarity limit?  
The well-known  black-disk limit for the scattering amplitude $f$ is reached when the maximal absorption, $h_{inel}=1/4$, takes place. It is definition of this limit, which  corresponds to the values $\mbox{Im}f =1/2$ and $\mbox{Re}f =0$ (cf. Eqs. (\ref{un}) and (\ref{re})).  
The impact-parameter analysis performed in \cite{alkin} implies that the black-disk limit has  been overcome at 
$\sqrt{s}=7$ TeV.  

In this comment we would like to point out that the most recent impact-parameter analysis performed by TOTEM at $\sqrt{s}=8$ TeV \cite{tot8} is in favor of this conclusion on exceeding the  black-disk limit. We do not discuss here the particular schemes of the elastic scattering amplitude unitarization, but it could be noted that the usual eikonal unitarization scheme meets problems with the black--disk limit exceeding.
Possible way to accommodate the situation is  consideration of the quasi-eikonal unitarization scheme \cite{sel, evg}.

Indeed, as it follows from Fig. 19 in \cite{tot8}, the value of $h_{el}$ at $b=0$ and $\sqrt{s}=8$ TeV is  ${\bf 0.31}$ while the black-disk limiting value is 
${\bf 0.25}$. It is true for the central $b$-dependence of  $h_{el}$ with maximum at $b=0$.  

Unfortunately, the impact parameter analysis performed in the paper \cite{tot8} does not account for  the respective experimental data error bars of the $d\sigma/dt$ measuremenrs, but, the analysis \cite{alkin} does and  its error bars  do not include the black disk limit at $b=0$.

As it was noted in \cite{tot8}, the peripheral dependence
 with maximal value of 0.05 at $b=1.2$ fm is also consistent with the data at this energy.
The existence of these two rather different forms, central and peripheral, is due to uncertainty in the nuclear phase  choice for the elastic scattering amplitude (cf. e.g. \cite{tot8}). The form and role of the nuclear phase is essential in the CNI (Coulomb-Nuclear Interference) region of very small values of $-t$.
However, the peripheral form is at variance with Regge and geometrical models for the elastic scattering.  

Moreover, the further observations can be made.
The slope parameter, an experimentally observed quantity, $B(s)$, \[B(s)\equiv \frac{d}{dt}\ln \frac{d\sigma}{dt}|_{-t=0},\]  is determined by the average value  $\langle b^2\rangle_{tot}$, where \cite{webb}:
\begin{equation}\label{btot}
\langle b^2\rangle_{tot}\sigma_{tot}(s)=\sum_{n=2}^{\infty}\langle b^2\rangle_{n}(s)\sigma_{n}(s),
\end{equation}
here $\sigma_{n}(s)$ is the $n$--particle production cross--section. This relation shows how the slope $B(s)$ is constructed from the individual elastic and inelastic contributions.
The explicit functional energy dependencies of the upper bounds for the functions
 \[ 
\langle b^2\rangle_{el,inel}(s)=\frac{\int_0^\infty b^3dbh_{el,inel}(s,b)}{\int_0^\infty bdbh_{el,inel}(s,b)}\]
have been obtained in
\cite{ajd}. They follow from the bound on $\langle b^2\rangle_{tot}$ and have the forms:
 \begin{equation}\label{bel}
\langle b^2\rangle_{el}(s)\leq  32\pi \frac{C^4}{\sigma_{el}(s)}\ln^4\frac{s}{s_0},
\end{equation}
\begin{equation}\label{binel}
\langle b^2\rangle_{inel}(s)\leq 8\pi \frac{C^4}{\sigma_{inel}(s)}\ln^4\frac{s}{s_0}.
\end{equation}
The two above bounds assume similar energy dependence in case when both $\sigma_{el}(s)$ and $\sigma_{inel}(s)$ 
have also similar $s$-dependence, say $\propto \ln^2 s$. It corresponds to the case of the black-disc limit saturation
and the upper bounds for $\langle b^2\rangle_{el}$ and $\langle b^2\rangle_{inel}$ are both proportional to $\ln^2 s$ under this scenario.

However, it is not the case when the elastic and inelastic cross-sections have different energy dependencies at $s\to\infty$, e.g. 
$\sigma_{el}\propto \ln^2 s$, while $\sigma_{inel}\propto \ln s$. Such dependencies of elastic and inelastic cross-sections
are typical for  the unitarity 
limit saturation when $\mbox{Im} f\to 1$ and $\mbox{Re} f \to 0$ at $s\to \infty$. In this case the upper bounds for 
$\langle b^2\rangle_{el}$ and $\langle b^2\rangle_{inel}$ would also have different energy dependencies, the former one
would be $\propto \ln^2 s$ while the latter one is proportional to $ \ln^3 s$. 
Such functional difference can be considered as an another  qualitative
issue in favor of a central form  for $h_{el}$ and a peripheral one for $h_{inel}$. 

It is worth to note here that the above two scenarios, corresponding to the  black-disk limit or unitarity limit saturation have been discussed, in particular, in \cite{anis} and the new data seems to be helpful in their discriminating.

Conclusion on central or peripheral forms of the overlap functions  can also be made on the grounds of unitarity and analyticity of the scattering amplitude in the Lehmann--Martin ellipse.
It is a straightforward consequence of the above general properties of the elastic scattering that the ratio of the overlap functions $h_{el}(s,b)/h_{inel}(s,b)$ is decreasing with $b$ like a linear
exponent $\sim e^{-\mu b}$ at large values of $b$ and fixed high energy value. It follows from a similar decreasing behavior of an elastic scattering amplitude \cite{grib}. Thus,  rather exotic situation with a central form of $h_{inel}$ and a peripheral one of $h_{el}$ is at variance with the results of unitarity and analyticity (under a natural assumption of a monotonous dependence of the overlap functions in the region of large values of $b$). 

The existence of the more central character of elastic scattering compared to the impact parameter distribution of the total probability of the inelastic processes is known for a long time and is in agreement with CERN ISR data \cite{npwe}, in particular, it was predicted at small $b$  values at the LHC energies (cf. e.g. \cite{jenk,degr}). 

It should be noted that peripheral $b$--dependence of $h_{inel}$ with maximum at $b\neq 0$ (i.e. with the fall down at small $ b$) was  depicted and discussed in \cite{npwe} for the energy $\sqrt{s}=3$ TeV. Currently,  various interpretations (\cite{edg,alb,arr}) of this effect have been proposed.

So, reasonably leaving aside the  peripheral option, one can conclude that the two independent impact parameter analysis of the  TOTEM data \cite{alkin,tot8} at the LHC energies $\sqrt{s}=7$ TeV and $8$ TeV indicate an existence of the transition to the reflective scattering mode \cite{refl} relevant to the asymptotic picture.

\section*{Acknowledgements}
We are grateful to M. Deile, J. Kaspar, E. Martynov and V. Petrov for the interesting discussions and correspondence.

\end{document}